# Muon Collider Interaction Region Design*

Y.I. Alexahin, E. Gianfelice-Wendt, V. V. Kashikhin, N.V. Mokhov, A.V. Zlobin,
*FNAL, Batavia IL, 60510 USA*

V.Y. Alexakhin,
*JINR, Dubna, 141980 Russia*

Design of a muon collider interaction region (IR) presents a number of challenges arising from low $\beta^* <$ 1 cm, correspondingly large beta-function values and beam sizes at IR magnets, as well as the necessity to protect superconducting magnets and collider detectors from muon decay products. As a consequence, the designs of the IR optics, magnets and machine-detector interface are strongly interlaced and iterative. A consistent solution for the 1.5 TeV c.o.m. muon collider IR is presented. It can provide an average luminosity of $10^{34}$ cm$^{-2}$s$^{-1}$ with an adequate protection of magnet and detector components.

PACS numbers 29.20.db, 84.71.Ba

## I. INTRODUCTION

A Muon Collider (MC) - proposed by G.I. Budker and A.N. Skrinsky more than 40 years ago [1] – has been extensively studied in U.S. during the past two decades [2, 3]. It is now considered as the most exciting option for the energy frontier machine in the post-LHC era. It has a number of important advantages over its competitor $e^+e^-$ collider: potentially higher energy, better energy resolution, larger cross-section for scalar particle production, smaller footprint, etc. [4]. However, in order to achieve a competitive level of luminosity a number of demanding requirements to the collider optics and the IR hardware should be satisfied arising from short muon lifetime and from relatively large values of the transverse emittance and momentum spread in muon beams that can realistically be obtained with ionization cooling [3].

Challenging as they are, these requirements are aggravated by limitations on the magnet maximum operating fields as well as by the necessity to protect superconducting magnets and collider detectors from muon decay products [5]. Therefore a holistic approach to the IR design should be developed tying together optics, magnet and shielding considerations.

The result of such an approach to the IR design of a muon collider with 1.5 TeV center of mass energy and an average luminosity of $10^{34}$ cm$^{-2}$s$^{-1}$ is presented in this paper. The particular value of the collision energy was chosen based on expectations of new physics at 1 TeV, though the future LHC results may point to a higher energy.

## II. IR LATTICE

The major problem to solve is correction of the IR quadrupoles chromaticity in such a way that the dynamic aperture remained sufficiently large and did not suffer much from strong beam-beam effects.

To achieve these goals a solution was proposed in the past based on special Chromatic Correction Sections (CCS) with compensated spherical aberrations [6]. Each CCS includes two sextupoles separated by a –I transformation so that their nonlinear kicks cancel out. There is an independent CCS for each transverse plane making the total of four chromaticity correction sextupoles on each side of the IP.

This approach has led to a number of muon collider designs, the best performance was demonstrated by a 4 TeV c.o.m. collider design by K. Oide [7]. According to it the vertical $\beta$-function in the final focus (FF) triplet is much larger than the horizontal one (up to 900 km for $\beta^* =$3 mm) and its chromatic perturbation is corrected first by a CCS starting at 180° vertical phase advance from the source (FF quads). However, very large $\beta$-function values together with large overall phase advance make the optics too sensitive to magnet field errors and misalignments.

### A. Chromatic Correction Scheme

In order to clarify the principle of the proposed scheme in this paper let us first recall the definition of the Montague chromatic functions [8]:

$$A_z = \frac{\partial}{\partial \delta}\alpha_z - \alpha_z B_z, \quad B_z = \frac{1}{\beta_z}\frac{\partial}{\partial \delta}\beta_z, \quad (1)$$
$$W_z = \sqrt{A_z^2 + B_z^2}, \quad \delta = \Delta p / p, \quad z = x, y$$

The form of equations which these functions obey depends on the set of dynamic variables used. With the choice of (non-canonical) pairs $(z, z')$ these equations are

$$A_z' = 2\varphi_z' B_z - \beta_z k, \quad B_z' = -2\varphi_z' A_z, \quad (2)$$

where $\varphi_z$ is the betatron phase advance, $k=\pm(K_1-D_x K_2)$ for $z=x,y$, $K_1$ and $K_2$ are normalized by $B\rho$ quadrupole and sextupole gradients, the prime denotes differentiation by path length.

Equations (2) show that initially only the Twiss $\alpha$-functions are perturbed, but as the betatron phase advance increases this initial perturbation – if left uncompensated – will be converted into a more dangerous perturbation of β-functions. Not to allow this to happen the correction sextupole must be placed at the same phase advance as the quadrupoles.

___________
* Work supported by Fermi Research Alliance, LLC under Contract DE-AC02-07CH11359 with the U.S. DOE.

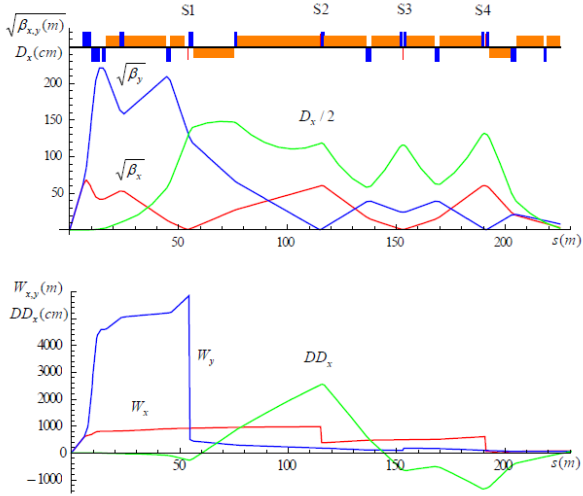

FIG. 1 (color). IR layout and optics functions (top) and chromatic functions (botom).

Figure 1 presents the IR layout which realizes this idea for the vertical plane, the horizontal chromatic function is much smaller (see Fig.1 lower plot) and can be corrected farther from the IP. Dipoles (shown at the top as orange rectangles) are placed next to the FF quadrupoles (blue rectangles) and generate a sufficiently large dispersion function at the S1 sextupole location. To increase dispersion the quadrupoles are displaced by ~1/10 aperture providing up to 2T bending field The lattice is symmetric with respect to the IP so that only the right half is shown.

Another principal difference of the proposed design is that we avoid using an error-prone CCS for the vertical plane relying only on smallness of the horizontal β-function at the S1 sextupole location: both resonance driving terms and detuning coefficients produced by a normal sextupole contain powers of $\beta_x$ and can be reduced with its help.

Such a recipe does not work for the horizontal plane: smallness of $\beta_y$ at a normal sextupole location is beneficial but does not suppress horizontal aberrations, so a CCS is still necessary with –I separated sextupole pair (marked as S2 and S4 in Fig. 1). Thus there is total of three sextupoles on each side of the IP for the Montague chromatic functions correction.

Correction of these functions – which is important by itself – also reduces the higher order chromaticity, i.e. the nonlinear dependence of betatron tunes on momentum. For the second order chromaticity we have [9]

$$\chi_z^{(2)} = \frac{1}{8\pi} \int_0^C (-kB_z \pm 2K_2 \frac{dD_x}{d\delta})\beta_z ds - \chi_z^{(1)} \quad (3)$$

with $\chi_z^{(1)}$ being the linear chromaticity, $z=x,y$.

TABLE I. Baseline muon collider parameters [10].

| Parameter | Unit | Value |
|---|---|---|
| Beam energy | TeV | 0.75 |
| Repetition rate | Hz | 15 |
| Average luminosity / IP | $10^{34}$/cm²/s | 1.1 |
| Number of IPs, $N_{IP}$ | - | 2 |
| Circumference, $C$ | km | 2.73 |
| $\beta^*$ | cm | 1 (0.5-2) |
| Momentum compaction, $\alpha_p$ | $10^{-5}$ | -1.3 |
| Normalized r.m.s. emittance, $\varepsilon_{\perp N}$ | π·mm·mrad | 25 |
| Momentum spread, $\sigma_p/p$ | % | 0.1 |
| Bunch length, $\sigma_s$ | cm | 1 |
| Number of muons / bunch | $10^{12}$ | 2 |
| Beam-beam parameter / IP, $\xi$ | - | 0.09 |
| RF voltage at 800 MHz | MV | 16 |

Equation (3) shows that the second order dispersion, $dD_x/d\delta$, also needs to be corrected. This is achieved by adjusting the relative values of the first order dispersion at sextupoles S2 and S4 and by installing an additional sextupole, S3, at the center of the horizontal CCS (Fig. 1).

This additional sextupole signifies the final departure from the concept of non-interleaved sextupole families which has also been abandoned in the design of the bending arcs [10].

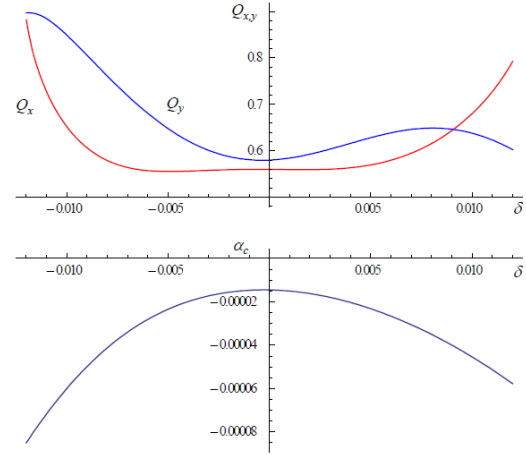

FIG. 2 (color). Fractional betatron tunes (top) and momentum compaction factor (botom) vs. momentum.

## B. Lattice Performance

Basic parameters of the muon beams and the collider lattice are given in Table 1. With relatively large emittances expected from the cooling channel and short bunch length the r.m.s. energy spread reaches 0.1% so that a momentum acceptance of at least ±0.3% is required.

Figure 2 shows the dependence on momentum of betatron tunes and momentum compaction factor obtained with some help from additional octupole and decapole correctors placed in the CCS. The stability range of ±1.2% significantly exceeds the minimum requirement.

Problems with the dynamic aperture (DA) and beam-beam effect in a muon collider are significantly alleviated by the fact that muons will be dumped after less than 2000 turns (see Section IV). In the result the high order resonances have little chance to show up. Preliminary studies [10] using MAD code demonstrated a good dynamic aperture (~$5\sigma$) in absence of magnet imperfections and beam-beam effect and only a modest DA reduction with the beam-beam parameter as large as 0.09 per IP[*].

The presented design raises a number of questions: large values of vertical $\beta$-function and therefore of the vertical beam-size in the IR quads and dipoles make it necessary to reconsider earlier magnet designs, closeness of the dipoles to IP may complicate the detector protection from γ-radiation emitted by decay electrons and positrons and from these electrons and positrons themselves.

These issues as well as problems with heat deposition in the magnet coils are considered in the subsequent sections.

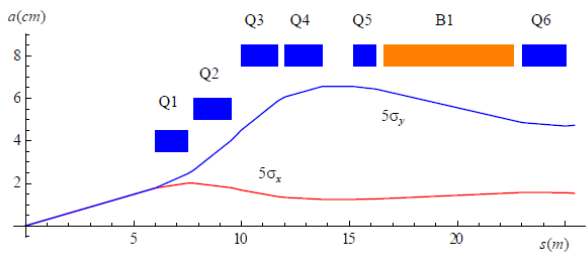

FIG. 3. Beam sizes and aperture of the FF magnets.

## III. IR MAGNET DESIGN

Figure 3 shows vertical and horizontal sizes of the muon beam corresponding to parameters from Table 1 and the inner radii of closest to IP magnets determined by the requirement $a > 5\sigma_{max}+1$ cm. A $5\sigma$ aperture radius may seem too small compared to $9\sigma_{max}$ aperture adopted for the LHC IR upgrade [11]. However, one should keep in mind that in MC there is no crossing angle and, due to short time the muons spend in the collider, there will be practically no diffusion so that the beams can be collimated at less than $4\sigma$ amplitudes; the remainder providing room for possible closed orbit excursions. In the actual magnet design, the bore radius was increased by additional 5 mm to provide more space for the beam pipe and annular helium channel.

---

[*] It should be noted that such values of beam-beam parameter were already achieved in $e+e-$ machines.

The expected level of magnetic fields in IR magnets suggests using $Nb_3Sn$ superconductor. This superconductor has the most appropriate combination of the critical parameters including the critical current density $J_c$, the critical temperature $T_c$, and the upper critical magnetic field $B_{c2}$ [12]. Cu-stabilized multi-filament $Nb_3Sn$ strands with $J_c$(12T, 4.2K)~3000 A/mm$^2$, strand diameter 0.7-1.0 mm and Cu/nonCu ratio~0.9-1.1 are commercially produced at the present time by industry in long length [13].

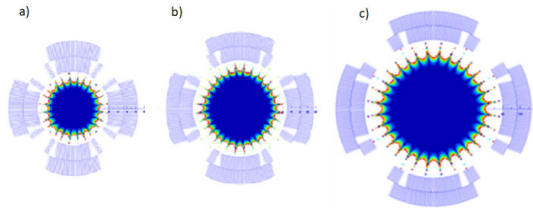

FIG. 4 (color). Cross-sections and a good-field region of Q1 (a), Q2 (b) and Q3-Q5 (c) quadrupoles. The dark blue color corresponds to the field error $|\delta B/B|<10^{-4}$.

### A. IR Quadrupoles

The IR doublets are made of relatively short quadrupoles (no more than 2 m long) to optimize their aperture according to the beam size variation and allow for placement of protecting tungsten masks between them. The first two quadrupoles in Fig. 3 are focusing ones and the next three are defocusing ones. The space between the 4$^{th}$ and 5$^{th}$ quadrupoles is reserved for beam diagnostics and correctors.

The cross-sections of MC IR quadrupoles based on two-layer shell-type $Nb_3Sn$ coils and cold iron yokes are shown in Fig. 4. Their parameters are summarized in Table 2. All the designs use wide 16.3 mm wide cable made of 37 strands 0.8 mm in diameter. Strand $J_c$(12T, 4.2K) after cabling is 2750 A/mm$^2$ and Cu/nonCu ratio is 1.17 [14]. To maximize the iron contribution to the quadrupole field gradient, it is separated from the coils by thin 10 mm spacers. The two-layer coil design and the total coil width were selected based on the results of $Nb_3Sn$ cable and coil R&D.

TABLE II. IR quadrupole parameters.

| Parameter | Unit | Q1 | Q2 | Q3 |
|---|---|---|---|---|
| Coil aperture | mm | 80 | 110 | 160 |
| Nominal gradient | T/m | 250 | 187 | -130 |
| Nominal current | kA | 16.61 | 15.3 | 14.2 |
| Quench gradient @ 4.5 K | T/m | 281.5 | 209.0 | 146.0 |
| Quench gradient @ 1.9 K | T/m | 307.6 | 228.4 | 159.5 |
| Coil quench field @ 4.5 K | T | 12.8 | 13.2 | 13.4 |
| Coil quench field @ 1.9 K | T | 14.0 | 14.4 | 14.8 |
| Magnetic length | m | 1.5 | 1.7 | 1.7 |

The nominal field in magnet coils is ~11-12 T whereas the maximum field is reaching ~13-15 T. As can be seen, all magnets have ~12% margin at 4.5 K, which is sufficient for the stable operation with the average heat deposition in the magnet mid-planes up to 1.7 mW/g. Operation at 1.9 K would increase the magnet margin to ~22% and their quench limit by a factor of 4.

The quench gradient and respectively operation margin of the IR quadrupoles at 4.5 K can be slightly increased if necessary by using wider (for example, 3- or 4-layers) and thus more complicate coils.

Geometrical field harmonics for IR quadrupoles Q1-Q5 are presented in Table 3.

Table III: Geometrical Harmonics at $R_{ref}$ ($10^{-4}$).

| Harmonic # | Q1 | Q2 | Q3 |
|---|---|---|---|
| $R_{ref}$ (mm) | 27 | 37 | 53 |
| $b_6$ | 0.000 | 0.000 | 0.000 |
| $b_{10}$ | -0.034 | 0.002 | 0.002 |
| $b_{14}$ | 0.862 | 0.090 | 0.086 |

The accelerator field quality is achieved within the circles (blue areas in Fig. 4) equal to 2/3 of the corresponding coil aperture. Saturation of the iron yoke and magnetization of cable and coil components and coil support structure will contribute to $b_6$. However, due to the fact that these magnets will operate at a constant field gradient all these components can be easily compensated by appropriate tuning the quadrupole coil geometry.

The designs and parameters (mainly high operating field and large operating margin) of the MC FF quadrupoles are quite challenging and thus need to be practically demonstrated. Since they are close to the parameters of quadrupoles being developed by US-LARP collaboration for the LHC luminosity upgrade [15, 16], the results of LARP magnet R&D will be applicable to the MC IR quadrupoles.

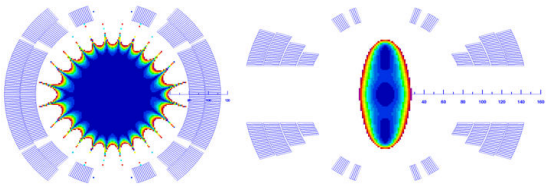

FIG. 5 (color). Cross-sections and a good-field region of the dipole B1 based on cos θ (left) and open mid-plane (right) coil design. The dark blue color corresponds to the field error of $|\delta B/B|<10^{-4}$.

### B. IR dipoles

The vertical elongation of the beam makes requirements to the IR dipoles quite different from those to the arc dipoles where the horizontal aperture must be larger due to the orbit sagitta and large dispersion contribution to the beam size. This allows using the traditional large-aperture cos θ design with a sufficiently thick inner tungsten liner to protect the cold mass from the muon decay products.

An alternative approach is the open mid-plane design concept, proposed for the MC SR dipoles [17], which allows the decay electrons to pass between the superconducting coils and be absorbed in high-Z rods cooled at liquid nitrogen or possibly at room temperatures and placed far from the coils. This reduces heat deposition in the coils and – potentially – background fluxes in the central tracker of the detector.

TABLE IV. IR dipole parameters.

| Parameter | Unit | Cosθ | Open midplane |
|---|---|---|---|
| Coil aperture | mm | 160 | 160 |
| Gap | mm | 0 | 55 |
| Nominal field | T | 8 | 8 |
| Nominal current | kA | 8.28 | 17.85 |
| Quench field @ 4.5 K | T | 12.46 | 9.82 |
| Magnetic length | m | 6 | 6 |

To remove 95% of radiation the full gap between the poles should be at least $5\sigma_y$ or 6 cm. This large gap limits the bending field which can be achieved with $Nb_3Sn$ coils and make it more difficult to attain an acceptable field quality in the required aperture.

Several options were considered for an open mid-plane dipole based on $Nb_3Sn$ superconductor with the required bending field of 8 T, good field quality in the aperture with 100 mm in vertical direction and 50 mm in horizontal direction, and appropriate margin at 4.5 K. The cross-sections of two-layer cosθ dipole design and most viable four-layer open midplane dipole design are shown in Fig. 5. The main parameters of cosθ and open midplane dipoles are reported in Table 4. Both dipole designs are based on 14.7 mm wide cable with 28 strands 1.0 mm in diameter [14]. Strand $J_c(12T,4.2K)=2750$ A/mm$^2$ includes possible ~10% cabling degradation and Cu/nonCu ratio is 1.0.

Table V: Geometrical Harmonics ($10^{-4}$).

| Harmonic # | Cosθ | Open midplane |
|---|---|---|
| $R_{ref}$ (mm) | 53 | 40 |
| $b_3$ | 0.04 | -5.88 |
| $b_5$ | 0.03 | -18.32 |
| $b_7$ | 0.40 | -17.11 |
| $b_9$ | 0.60 | -4.61 |

Geometrical field harmonics at the corresponding reference radii for IR dipoles B1 based on two alternative magnet designs are presented in Table 5. In the traditional cosθ design the good field quality is provided within the circle with a radius of 60 mm (blue area in Fig. 5a). In the open midplane design the accelerator field quality is provided within a required elliptical area with 50 mm horizontal and 110 mm vertical size (blue area in Fig. 5b). In this design it was achieved by an appropriate combination of relatively large values of low-order

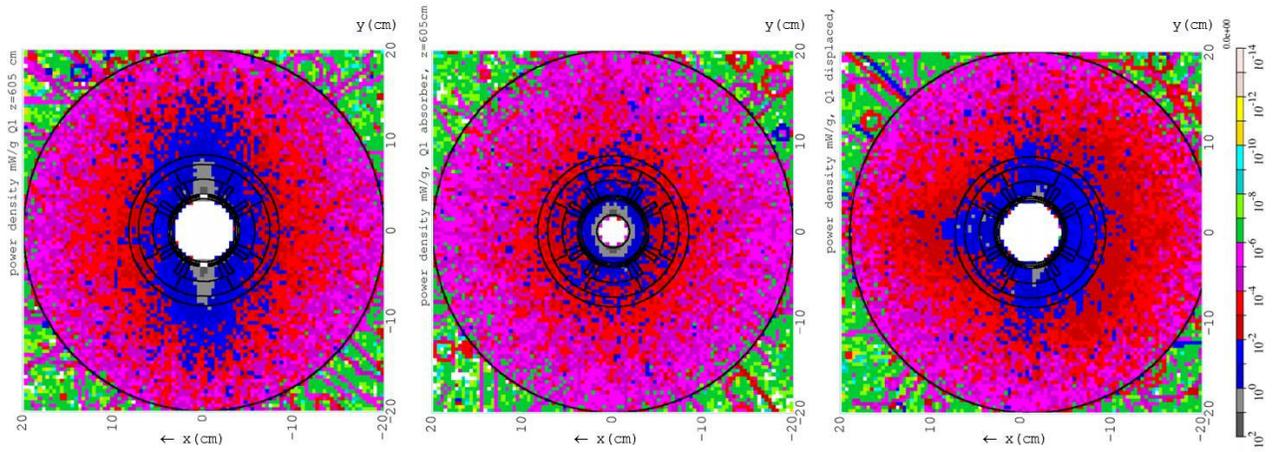

FIG. 6 (color). Deposited power density in Q1 (mW/g) for three cases: "standard" (left), with absorbers inside (center) and with horizontal displacement (right). Larger radii are on the left of the plots.

geometrical harmonics. As in the case of IR quadrupoles, the saturation of iron yoke and the magnetization of cable and coil components and coil support structure will contribute to the low order field harmonics, mainly to $b_3$ and $b_5$. All these contributions will be compensated by re-optimizing the low order harmonics at the operating field.

As it follows from Table 4, the traditional $\cos\theta$ design provides larger maximum field and respectively larger operation margin than the open mid-plane design. It is also more straightforward from the viewpoint of fabrication and cold mass cooling. However, the aperture of this magnet, the coil volume and the Lorentz force level depend on the absorber size which make this design also quite challenging. Both designs require significant R&D efforts.

## IV. ENERGY DEPOSITION IN MAGNETS

Energy deposition and detector backgrounds are simulated with the MARS15 code [18]. All the related details of geometry, materials distributions and magnetic fields are implemented into the model for lattice elements and tunnel in the ±200-m region from IP, detector components [19], experimental hall and machine-detector interface. To protect SC magnets and detector, tungsten masks in the interconnect regions, liners in magnet apertures (wherever needed), and a sophisticated tungsten cone inside the detector [5] were implemented into the model and carefully optimized. The muon beam with parameters cited in Table 1 was assumed to be aborted after 1500 turns when the luminosity is reduced by a factor of ~6.

Three cases were considered: (i) "standard" when 10-cm long tungsten masks with 5 $\sigma_{x,y}$ elliptic openings are put in the IR magnet interconnect regions; (ii) with additional tungsten liners inside the quadrupoles leaving a 5 $\sigma_{x,y}$ elliptic aperture for the beam; (iii) as first case, but with the IR quadrupoles displaced horizontally by 0.1 of their apertures, so as to provide ~2 T bending field. This additional field helps also facilitate chromaticity correction by increasing dispersion at the sextupoles, and deflect low-energy charged particles from the detector.

Power density isocontours at shower maximum in the first quadrupole are shown in Fig. 6, while Fig. 7 displays such profiles in the IR dipole B1. Maximum values of power density in the most vulnerable magnets are presented in Table 6. One can see that quadrupole displacement reduces power density but not enough to avoid using liners inside quadrupoles. Combining all the three cases has a potential of keeping peak power density in the IR magnets below the quench limits of about 5 mW/g with a necessary safety margin (typically a factor of three).

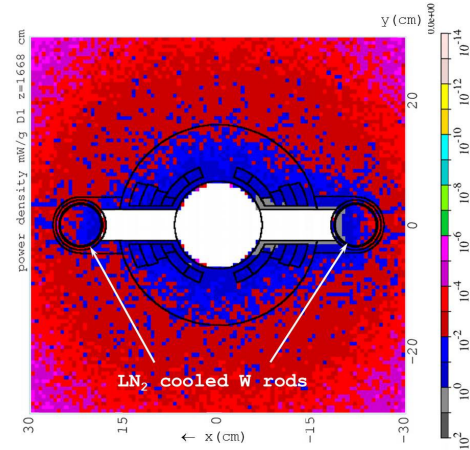

FIG. 7 (color). Power density (mW/g) in B1 dipole for case (iii).

TABLE VI. Peak power density (mW/g) in most vulnerable magnets in three considered cases.

| Magnet | (i) | (ii) | (iii) |
|---|---|---|---|
| Q1 | 5.0 | 1.0 | 3.0 |
| Q2 | 10. | 1.0 | 10. |
| Q5 | 3.7 | 2.0 | 3.7 |
| B1 | 3.0 | 2.6 | 1.9 |
| Q6 | 3.6 | 2.6 | 2.0 |

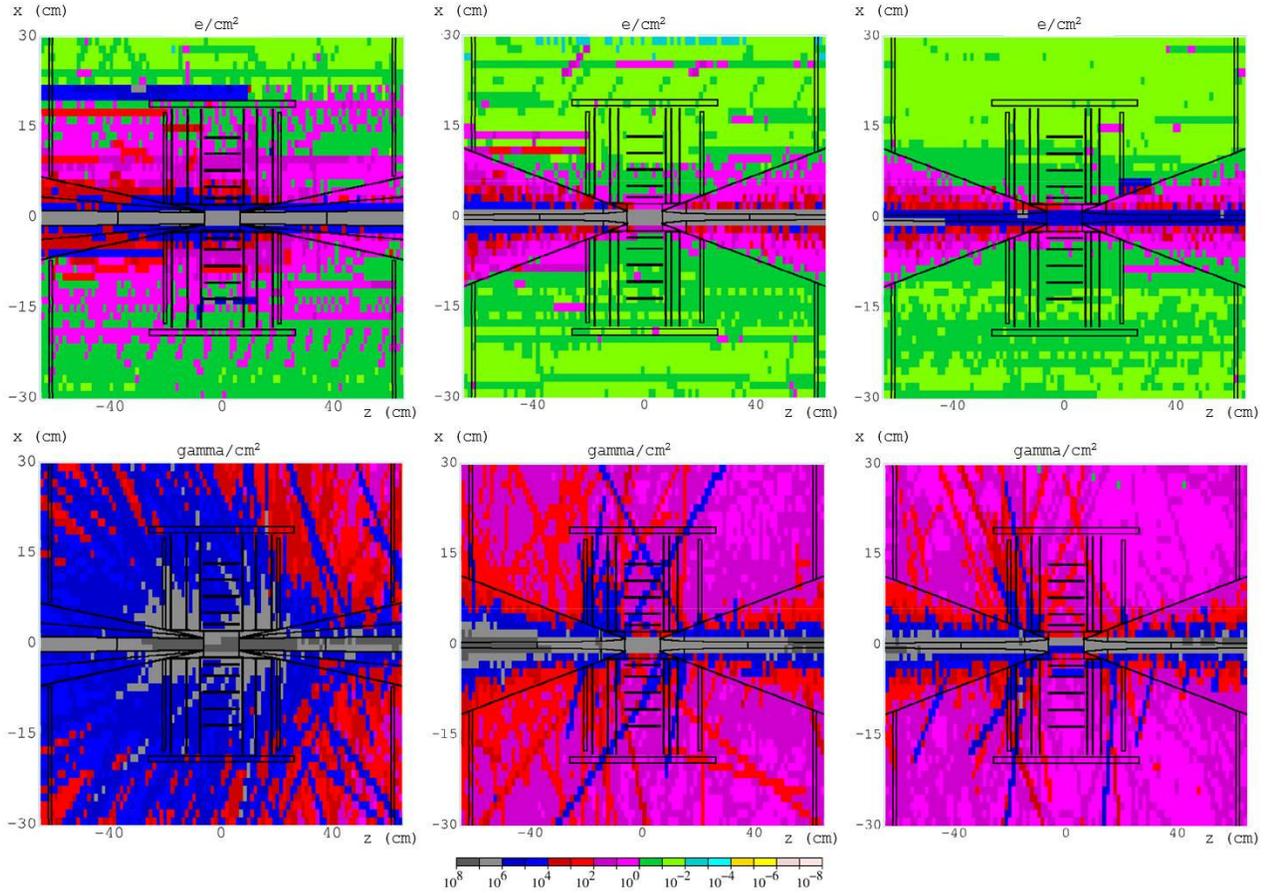

FIG. 8 (color). Electron (top) and gamma (bottom) fluxes in the detector in three cases described in the text.

## V. DETECTOR BACKGROUNDS

Figure 8 compares calculated electron and gamma fluxes for the following cases: left – no masks between magnets, 6° cone with a 5σ radius liner up to 2 m from IP; center - 5σ masks inserted between FF quads, cone angle increased to 10°, 5σ liner up to 1 m from IP; right – same as above plus FF quad displacement.

The masks and increased cone angle reduce the electron and gamma fluxes by factors 300 and 20, respectively. Displacing the FF quads slightly increases the electron flux (by up to 50%) but decreases the gamma flux by another factor of 15, so the overall effect of quad displacement may be considered as positive.

Results of further optimization of the cone nose geometry are presented in Fig. 9. It shows gamma flux as a function of the angle of inner cone opening towards IP at the outer cone angle of 10°. For such a cone and a set of other the most optimal parameters – as it is seen now – the maximum neutron fluence and absorbed dose in the innermost layer of the silicon tracker for a one-year operation are at a 10% level of that in the LHC detectors at the luminosity of $10^{34}$ cm$^{-1}$s$^{-1}$. Photon fluence is several times higher than that at the LHC.

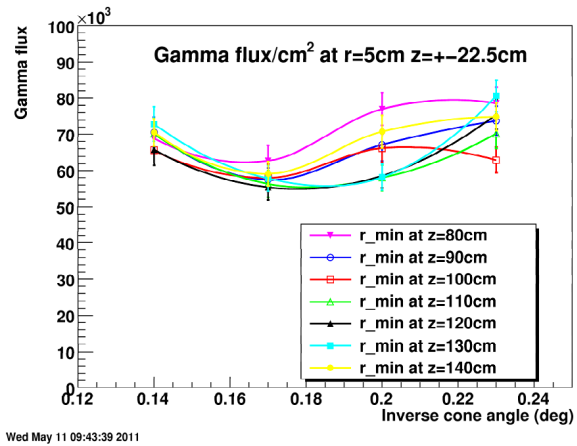

FIG. 9 (color). Gamma flux vs. inner cone angle at different positions of minimal aperture from IP

## VI. SUMMARY & OUTLOOK

The presented interaction region lattice is a part of the complete muon collider storage ring design which satisfies all requirements from the beam dynamics point of view in the considered case of 1.5 TeV center of mass energy and the average luminosity of $10^{34}$ cm$^{-2}$s$^{-1}$.

All the required IR magnets can be built using the Nb$_3$Sn technology which is being developed for the LHC

luminosity upgrade and shows very promising results for a Muon Collider Storage Ring and Interaction Regions. Using a combination of special measures (internal absorbers and masks) the heat deposition in IR magnets can be reduced below the quench limit of $Nb_3Sn$ magnets at 4.5 K with a safety margin.

With the proposed protective measures implemented in the machine-detector interface, the calculated backgrounds are comparable to those expected at LHC for the same luminosity.

Further studies and optimization of the 1.5 TeV muon collider design need to be focused on:

- Feasibility studies and modelling of the open-midplane dipole design. The studies and development of large-aperture traditional dipole magnets with comparable operating parameters are supported by some other R&D programs (see, for example, [20]).
- Reduction of detector backgrounds by optimizing parameters of the protective cone and other machine-detector interface elements.
- Adding a collimation scheme to the muon collider lattice design, which actually should be extraction of the beam halo [21].

At the same time the work on a more challenging muon collider design with a 3-TeV center-of-mass energy has also been started. It will require even stronger SC magnets and will have to address such issues as the neutrino-induced radiation [22].

## ACKNOWLEDGEMENTS

The authors are grateful to S. Geer, R. Palmer and A. Tollestrup for many useful remarks and to Dr. K. Oide for kindly providing the detail of his MC design.